\documentclass[final,5p,times,twocolumn]{elsarticle} 

\newcommand{\mpw}{RD50-MPW4\xspace}

\usepackage[hidelinks]{hyperref}
\usepackage{siunitx}
\usepackage{graphicx}
\usepackage{amsmath}   
\usepackage{subcaption}
\usepackage{multirow}
\usepackage{xspace}
 \usepackage{mhchem}
\usepackage[switch, modulo]{lineno}
\modulolinenumbers[1]

\begin{document}

\begin{frontmatter}
\title{Enhancing Radiation Hardness and Granularity in HV-CMOS: The RD50-MPW4 Sensor}

\author[1]{B. Pilsl\corref{cor1}}
\ead{Bernhard.Pilsl@oeaw.ac.at}

\author[1]{T. Bergauer}
\author[5]{R. Casanova}
\author[1]{H. Handerkas}
\author[1]{C. Irmler}
\author[3]{U. Kraemer}
\author[6]{R. Marco-Hernandez}
\author[6]{J. Mazorra de Cos}
\author[7]{F. R. Palomo}
\author[1]{S. Portschy}
\author[2]{S. Powell}
\author[1]{P. Sieberer\fnref{fn2}}
\author[3]{J. Sonneveld}
\author[1]{H. Steininger}
\author[2]{E. Vilella}
\author[2]{B. Wade}
\author[2]{C. Zhang}
\author[4]{S. Zhang}

\cortext[cor1]{Corresponding author}
\fntext[fn2]{Now at Paul Scherrer Institut (PSI).}

\affiliation[1]{organization={Austrian Academy of Sciences, Institute of High Energy Physics}, 
                 addressline={Nikolsdorfergasse 18},
                 postcode={1050}, 
                 city={Vienna}, 
                 country={Austria}}                 
\affiliation[2]{organization={Department of Physics, University of Liverpool},
                 addressline={Oliver Lodge Building, Oxford Street},
                 postcode={L69 7ZE}, 
                 city={Liverpool},
                 country={UK}}
\affiliation[3]{organization={NIKHEF},
                 addressline={Science Park 105},
                 postcode={1098 XG},
                 city={Amsterdam},
                 country={Netherlands}}
\affiliation[4]{organization={Physikalisches Institut, Rheinische Friedrich-Wilhelms-Universitaet Bonn},
                 addressline={Nussallee 12},
                 postcode={53115}, 
                 city={Bonn},
                 country={Germany}}
\affiliation[5]{organization={Institute for High Energy Physics (IFAE), 
Autonomous University of Barcelona (UAB)},
                 addressline={Bellaterra},
                 postcode={08193}, 
                 city={Barcelona},
                 country={Spain}}
\affiliation[7]{organization={Department of Electronic Engineering, University of Sevilla},
                 addressline={Calle San Fernando 4},
                 postcode={41092}, 
                 city={Sevilla},
                 country={Spain}}
\affiliation[6]{organization={Instituto de Fisica Corpuscular (IFIC), CSIC-UV},
                 addressline={Parque Cientifico, Catedratico Jose Beltran 2},
                 postcode={46980}, 
                 city={Paterna (Valencia)},
                 country={Spain}}



\begin{abstract}
The latest HV-CMOS pixel sensor developed by the former CERN-RD50-CMOS group, known as the \mpw, demonstrates competitive radiation tolerance, spatial granularity, and timing resolution -- key requirements for future high-energy physics experiments such as the HL-LHC and FCC. Fabricated using a \SI{150}{nm} CMOS process by \emph{LFoundry}, it introduces several improvements over its predecessor, the \emph{RD50-MPW3}, including separated power domains for reduced noise, a new backside biasing scheme, and an enhanced guard ring structure, enabling operation at bias voltages up to \SI{800}{V}.

Tests with non-irradiated samples achieved hit detection efficiencies exceeding \SI{99.9}{\%} and a spatial resolution around \SI{16}{\mu m}. Neutron-irradiated sensors were characterized using IV measurements and test-beam campaigns, confirming the sensor's robustness in high-radiation environments. The results highlight the ability of HV-CMOS technology to restore hit detection efficiency post-irradiation by increasing the applied bias voltage. Details of these measurements and timing performance are presented in this paper.
\end{abstract}

\end{frontmatter}

\section{Introduction}

The \mpw is an HV-CMOS DMAPS sensor fabricated by \emph{LFoundry} in a \SI{150}{nm} process utilizing a large collection electrode design. It comprises a $64 \times 64$ pixel matrix with a square pixel pitch of $62 \times \SI{62}{\mu m^2}$. Each hit has two 8-bit timestamps, one for the leading edge and one for the trailing edge of the comparator's output signal~\cite{Sieberer_2023}. Both timestamps are recorded based on a \SI{25}{ns} clock. This timing information enables the reconstruction of the time-over-threshold (ToT), which is proportional to the signal amplitude and can provide information about the deposited charge.

So far, the \mpw has presented itself as a highly efficient sensor with a good spatial resolution for a large collection electrode design~\cite{PilslPISA24}. The architecture allows for operation at high bias voltages, enhancing the depleted volume and ensuring fast and complete charge collection. These properties are particularly desirable in environments with high radiation fluences.

To investigate the post-irradiation performance, several sensor samples were irradiated with neutrons to various fluence levels as summarized in table~\ref{tab:samples}. Samples from wafer 3 (W3) received backside processing, in which they were thinned to $\SI{280}{\mu m}$ and metalized on the backside, enabling the connection of the bias voltage from the back. These sensors can still be biased from the topside, offering flexibility for experimental studies and system integration.

To enable a comparison between the two biasing configurations, most of the results presented in this paper include both top and backside biasing modes. This comparative approach helps to assess the potential benefits of backside biasing in mitigating radiation damage effects and optimizing electric field uniformity. This is particularly relevant because, for ensuring radiation hardness, a homogeneous electric field across the sensor volume is crucial — especially after irradiation when field distortions can significantly impact charge collection~\cite{Moll:1999kv}. The following sections detail the IV behavior, timing, and detection efficiency of the \mpw sensor under both non-irradiated and irradiated conditions, with a focus on understanding how these factors evolve with temperature and fluence. The sample irradiated to $\SI{1E15}{n_{eq}cm^{-2}}$ resides from W3, which did not get the additional backside processing and can, therefore, only be biased from the top.

\begin{table}[htbp]
\setlength{\tabcolsep}{4pt}
\renewcommand{\arraystretch}{0.8}
\centering
\small
\begin{tabular}{c|c}
Wafer & Fluence [\si{n_{\text{eq}} cm^{-2}}] \\ \hline
\multirow{3}{*}{W3} & \SI{1E14}{} \\
                    & \SI{1E15}{} \\
                    & \SI{1E16}{} \\ \hline
W8                  & \SI{1E15}{} \\
\end{tabular}
\caption{Samples irradiated to different fluence levels.}
\label{tab:samples}
\end{table}

\section{IV characteristics}
\begin{figure}[htbp]
\centering
\begin{subfigure}{0.4\textwidth}
\centering
\includegraphics[width=\textwidth]{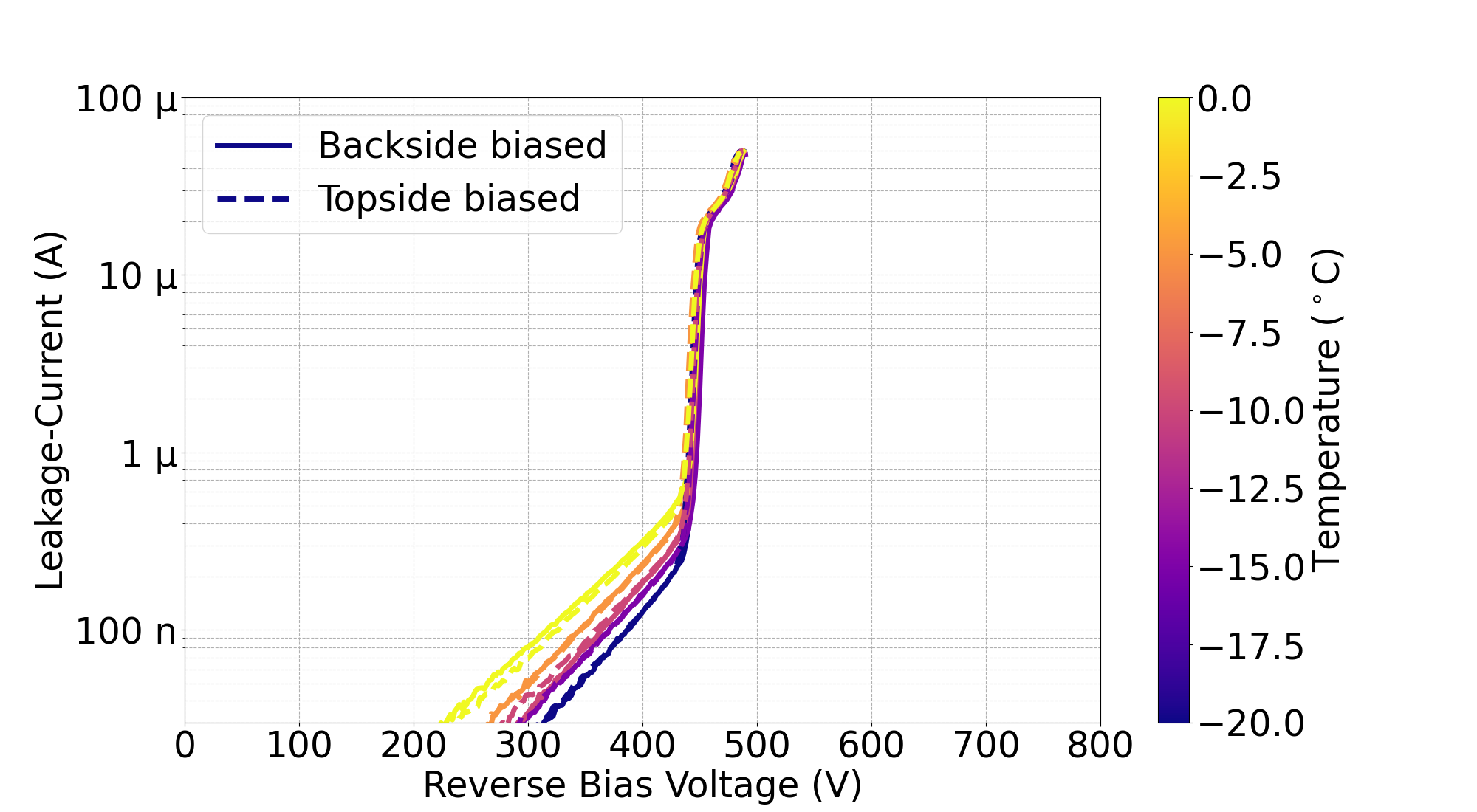}
\caption{Non irradiated sample.}
\end{subfigure}
\begin{subfigure}{0.4\textwidth}
\centering
\includegraphics[width=\textwidth]{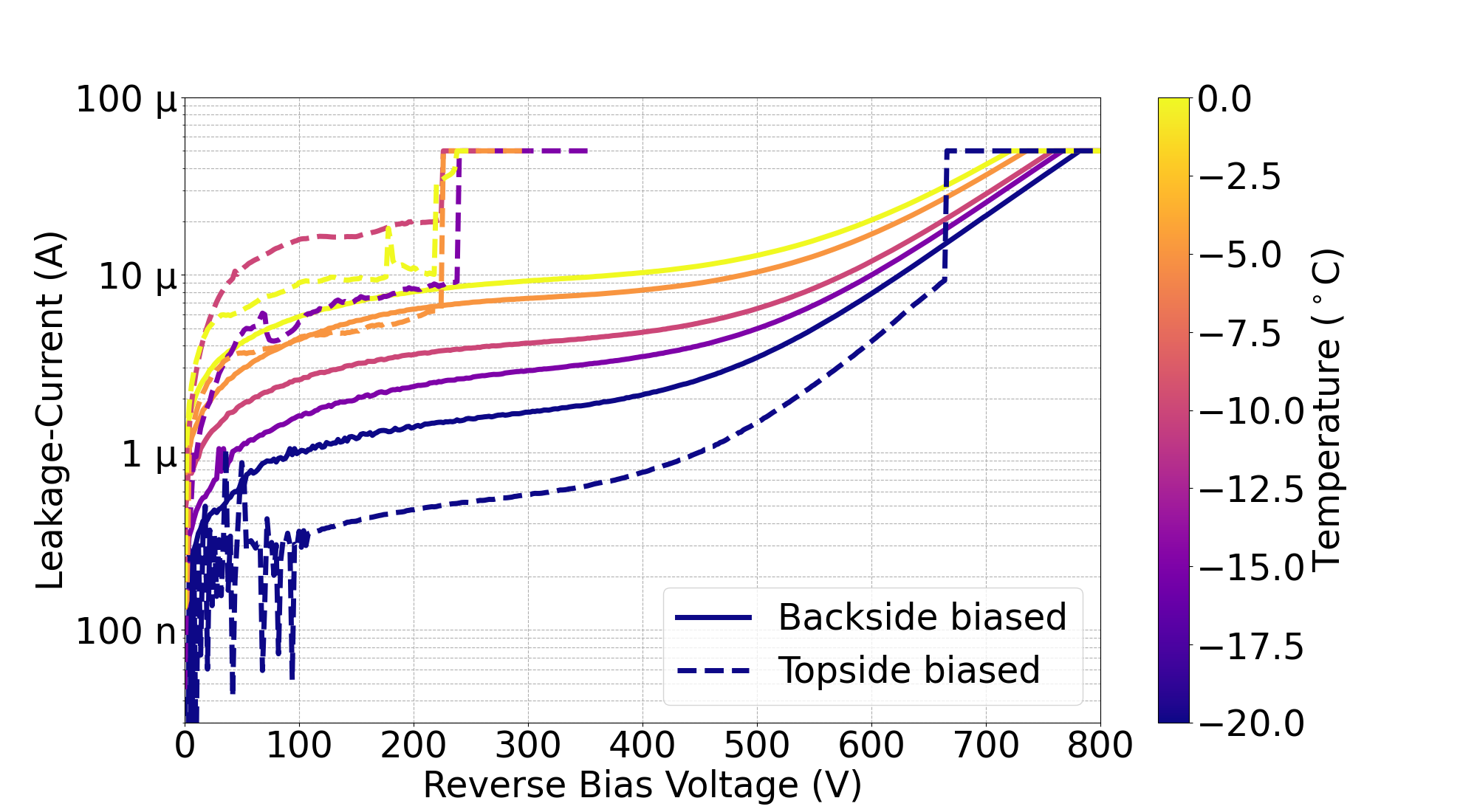}
\caption{Sample irradiated to $\SI{1E14}{n_{eq}cm^{-2}}$.}
\end{subfigure}
\begin{subfigure}{0.4\textwidth}
\centering
\includegraphics[width=\textwidth]{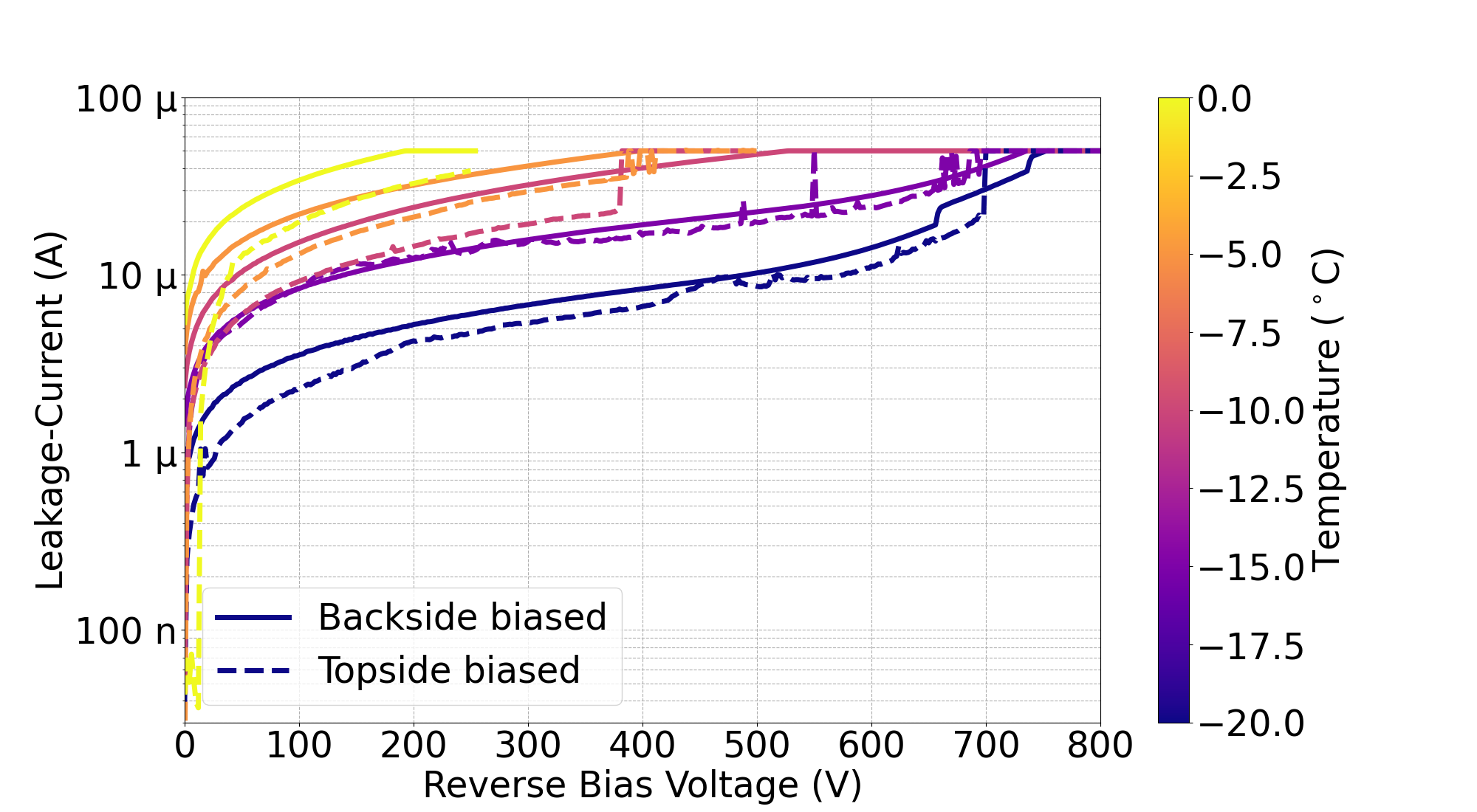}
\caption{Sample irradiated to $\SI{1E15}{n_{eq}cm^{-2}}$.}
\end{subfigure}
\begin{subfigure}{0.4\textwidth}
\centering
\includegraphics[width=\textwidth]{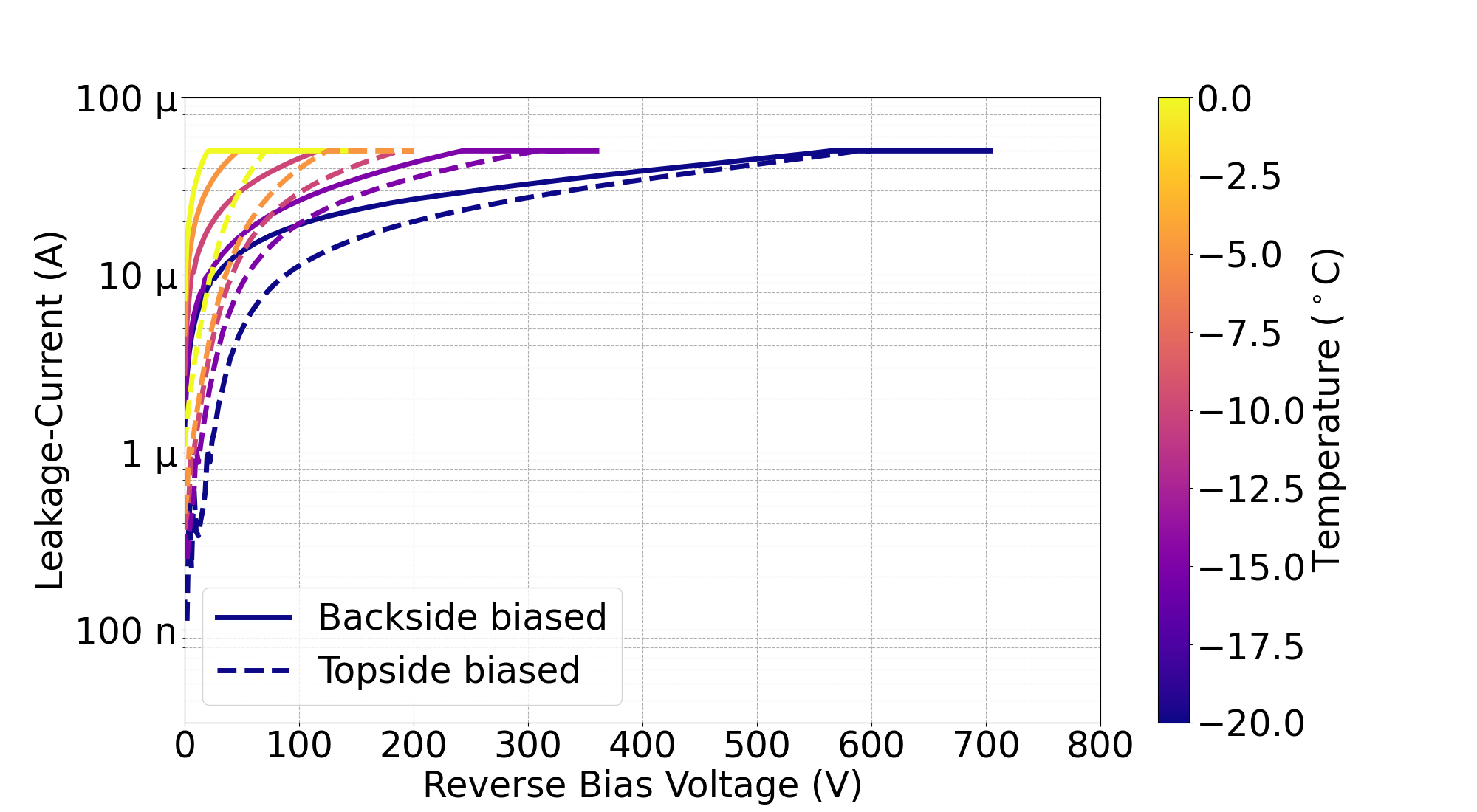}
\caption{Sample irradiated to $\SI{1E16}{n_{eq}cm^{-2}}$.}
\end{subfigure}
\caption{IV characteristics of the various irradiated samples at different temperatures. In total 4 different temperatures, in steps of $\SI{5}{\degree C}$ in the range from $[-20, 0] \SI{}{\degree C}$ were evaluated. The compliance current of the voltage source was set to $\SI{50}{\mu A}$ to ensure no damage to the samples. All samples are taken from W3.}
\label{fig:ivs}
\end{figure}

The full matrix was biased at different temperatures to study sensor performance under realistic conditions, with temperatures measured by a thermocouple adjacent to the sensor. These measurements are presented in figure~\ref{fig:ivs}, while test structure measurements are presented in~\cite{VilellaTWEPP24}. A strong temperature dependence of the leakage current is observed after irradiation. Lowering the temperature to $\SI{-20}{\degree C}$ allows the sample irradiated to $\SI{1E16}{n_{eq}cm^{-2}}$ to be biased at $\approx \SI{600}{V}$, which is approximately 8 times higher than the achievable bias voltage of $\approx \SI{80}{V}$ at  $\SI{0}{\degree C}$. A significant difference between back and topside biasing is observed at an irradiation level of $\SI{1E14}{n_{eq}cm^{-2}}$, whereas for the other investigated irradiation levels, the biasing scheme has no major impact on the leakage current. This supports the claim that at lower irradiation levels, surface effects such as oxide charge accumulation and interface states can significantly influence the electric field distribution during operation, depending on the applied biasing scheme. These surface-induced modifications to the field profile are more prominent when bulk damage is limited. However, at higher fluences, bulk damage dominates through the creation of deep-level traps and generation centers in the silicon lattice, which drive the leakage current and overshadow surface-related effects, making the biasing scheme less influential on the amount of leakage current~\cite{Lindstroem:2001, Hughes}.

\section{Test-Beam evaluation}

A test-beam campaign at the \emph{DESY II} beam facility~\cite{DIENER2019265} was conducted to assess the performance of irradiated samples and compare them with non-irradiated ones, providing insights into radiation damage effects on the \mpw. Six planes of the \emph{Adenium} telescope~\cite{Liu_2023} were used for reference. An additional plane of the \textsc{TelePix2}\cite{tp2Ref} served as a timing layer for track timestamping and as a region of interest (ROI) trigger. This ROI trigger got fed into the \emph{AIDA-2020-TLU}~\cite{AidaTLU}, enabling synchronization between the \mpw (DUT) and the telescope. While the telescope is synchronized via trigger numbers, the DUT uses timestamps from the leading edge and an overflow counter, with the first assigned on-chip and the latter one on the firmware level. A Peltier-element based cooling system was installed to facilitate cooling at the test-beam site, providing indirect cooling through a copper piece down to approximately $\SI{-15}{\degree C}$.

The data analysis utilized the \emph{Corryvreckan}~\cite{corry} framework.

\subsection{Timing Performance}

\begin{figure}[htbp]
\centering
\includegraphics[width=.39\textwidth]{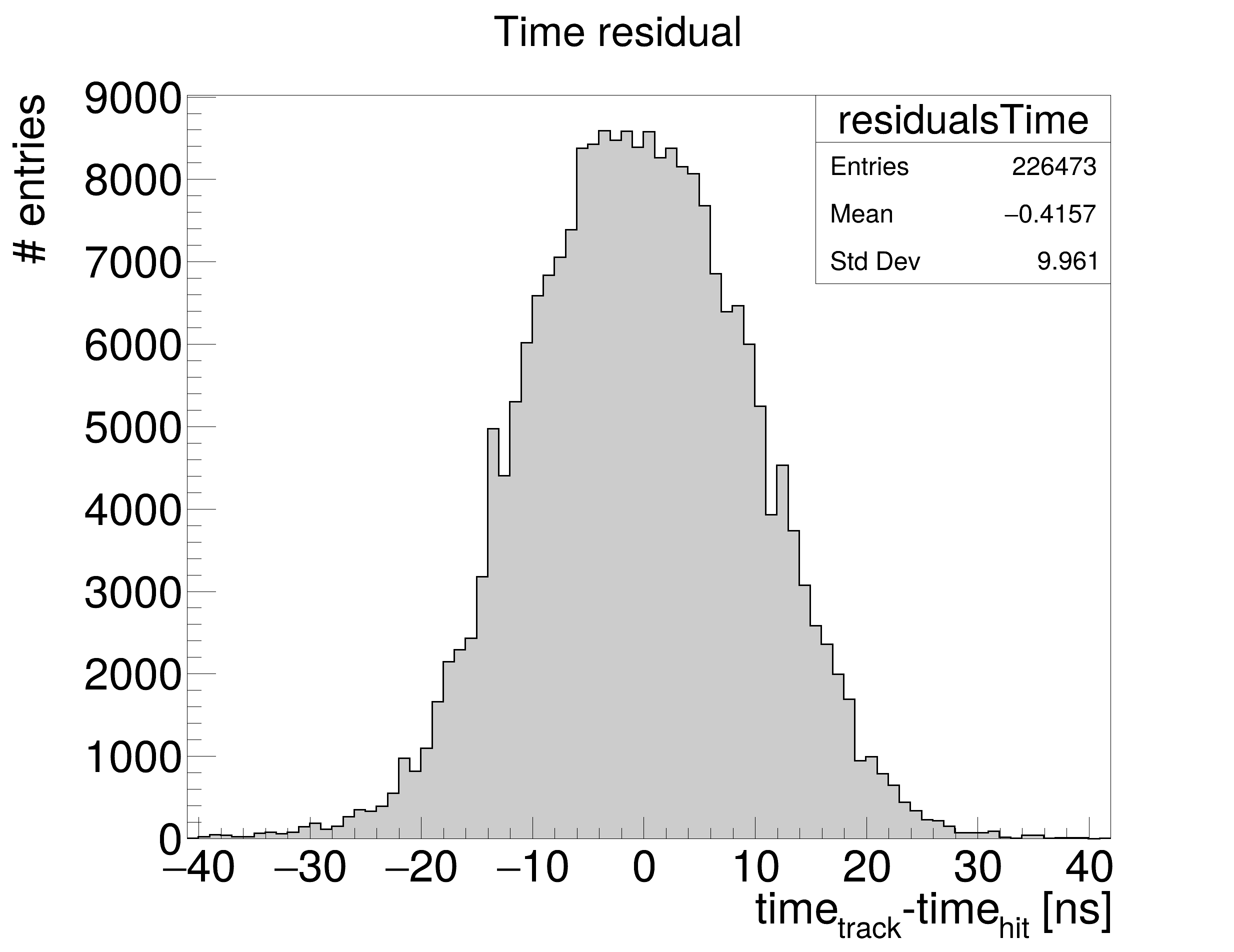}
\caption{Time residuals showing the main peak centered at $\Delta t \approx \SI{0}{ns}$.}
\label{fig:tRes}
\end{figure}

Calculating the time residuals by the difference of the track timestamp, assigned by the \textsc{TelePix2}, and the hit timestamp from the DUT, as shown in figure~\ref{fig:tRes}, allows to depict the timing resolution. Truncating the distribution to the innermost 99\% of the main peak and evaluating the standard deviation allows to depict a timing resolution of \SI{9.8}{ns}.

\begin{figure}[htbp]
\centering
	\includegraphics[width=.39\textwidth]{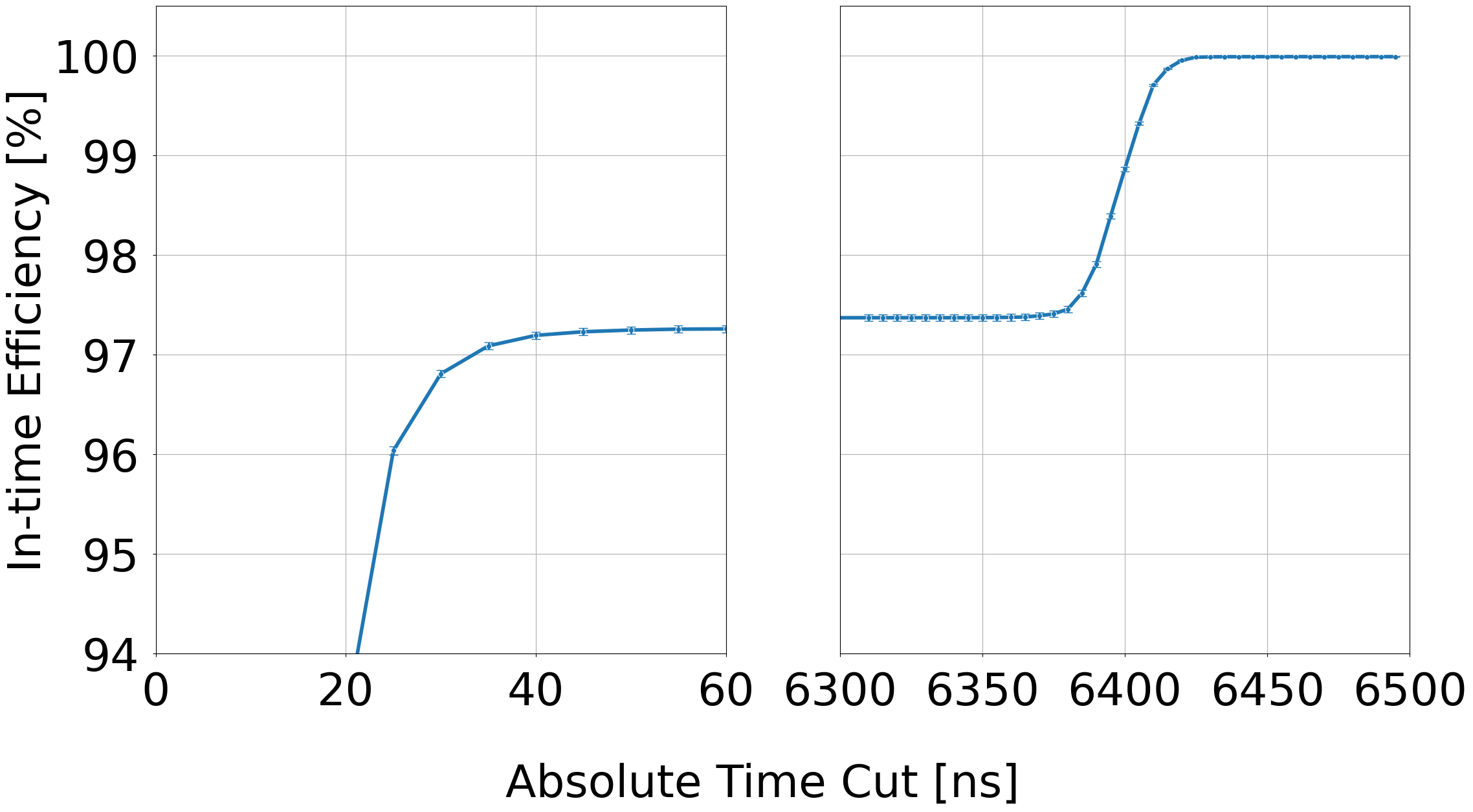}
	\caption{In time efficiency for various timing cuts of a non-irradiated sample. The cut range from $[\SI{60}{ns}, \SI{6.3}{\mu s}]$ is not shown as no major in-time efficiency increase is observed in this region.}
	\label{fig:inTimeEffi}
\end{figure}

The in-time efficiency is evaluated by allowing only hits within a defined time window relative to the track's timestamp. Figure~\ref{fig:inTimeEffi} shows the hit detection efficiency with different timing cuts applied. Two major in-time efficiency increase regions are observed. The first increase up to $\approx \SI{97}{\%}$ shows the efficiency losses when cutting on the central time residuals peak, while the second increase region, which contains the remaining $\approx \SI{3}{\%}$, corresponds to several hits which got a wrong overflow counter value assigned to and are thereby displaced by $256 \times \SI{25}{ns} = \SI{6.4}{\mu s}$. The DAQ system assigns the overflow counter to the various hits off-chip and can cause a jump in the time domain depending on the hit rate. Within the standard LHC bunch crossing time window of \SI{25}{ns}, an in-time hit detection efficiency $> \SI{96}{\%}$ is achieved.

\subsection{Direct comparison of fluence levels}

\begin{table}[htbp]
\small
\begin{tabular}{c|c|c|c}
Fluence [$\SI{}{n_{eq}cm^{-2}}$] & $\text{Cluster-size}$ & $\text{ToT} [\SI{25}{ns}]$ & Efficiency [\%] \\ \hline
Non-Irradiated & $1.14\pm 0.43$ & $7.95 \pm 3.16$ & 99.8\\
\SI{1E14}{} & $1.15 \pm 0.46$ & $13.55 \pm 5.79$ & 99.5 \\
\SI{3E14}{} & $1.12 \pm 0.42$ & $4.99 \pm 2.04$ & 85.5 \\
\SI{1E15}{} & $1.04\pm 0.21$ & $2.07 \pm 1.09$ & 9.2
\end{tabular}
\caption{Arithmetic mean and and standard deviation of the cluster size, the ToT and the hit detection efficiency at common bias voltage and threshold settings for different irradiation levels. The low efficiency values are a result of the high threshold and the relatively low bias voltage.}
\label{tab:directComp}
\end{table}

\begin{figure}[htbp]
    \centering
\begin{subfigure}{0.23\textwidth}
    \centering
\includegraphics[width=\textwidth]{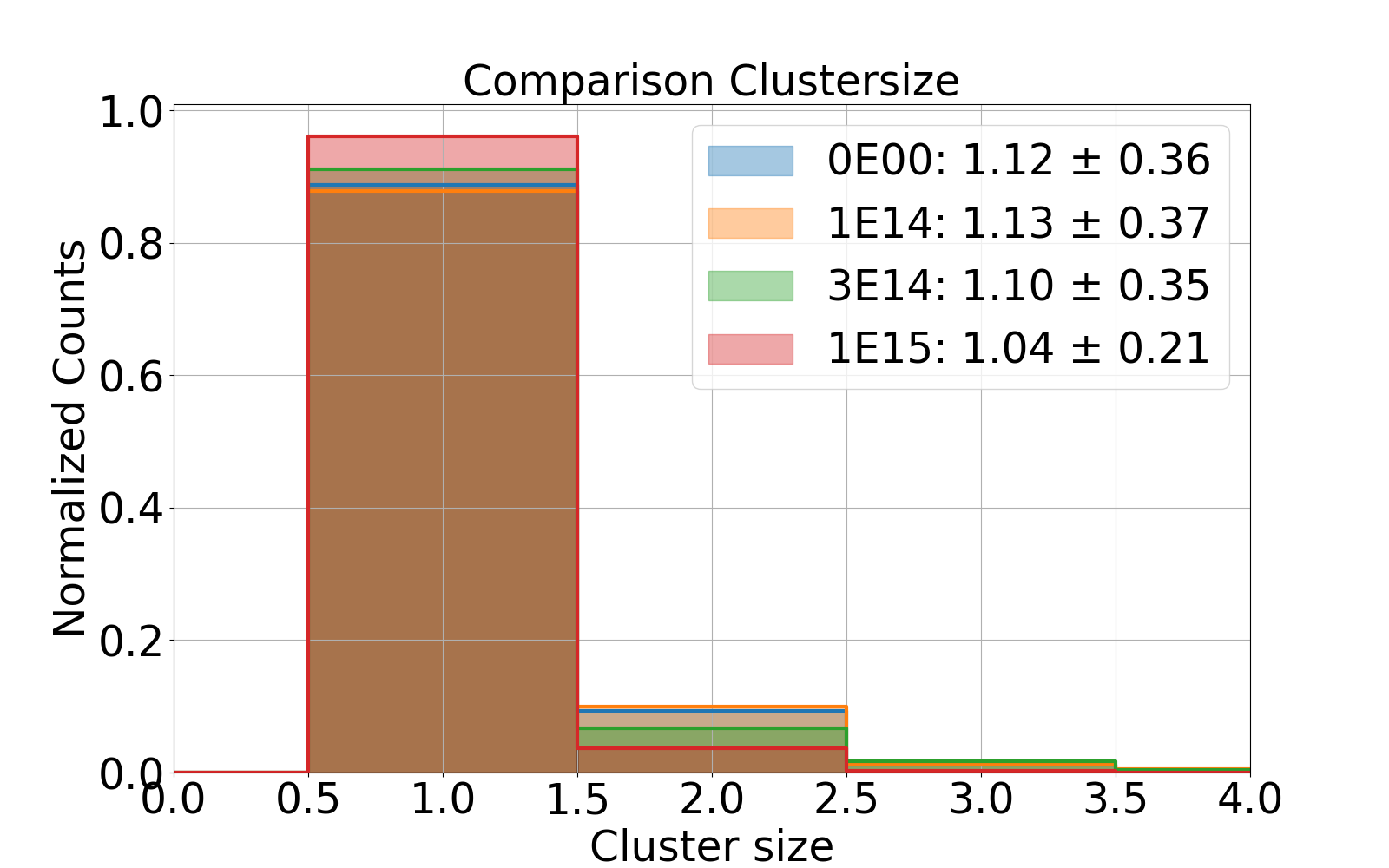}
\caption{Comparison of the normalized cluster size.}
\label{fig:directCompClstr}
\end{subfigure}
\begin{subfigure}{0.23\textwidth}
    \centering
\includegraphics[width=\textwidth]{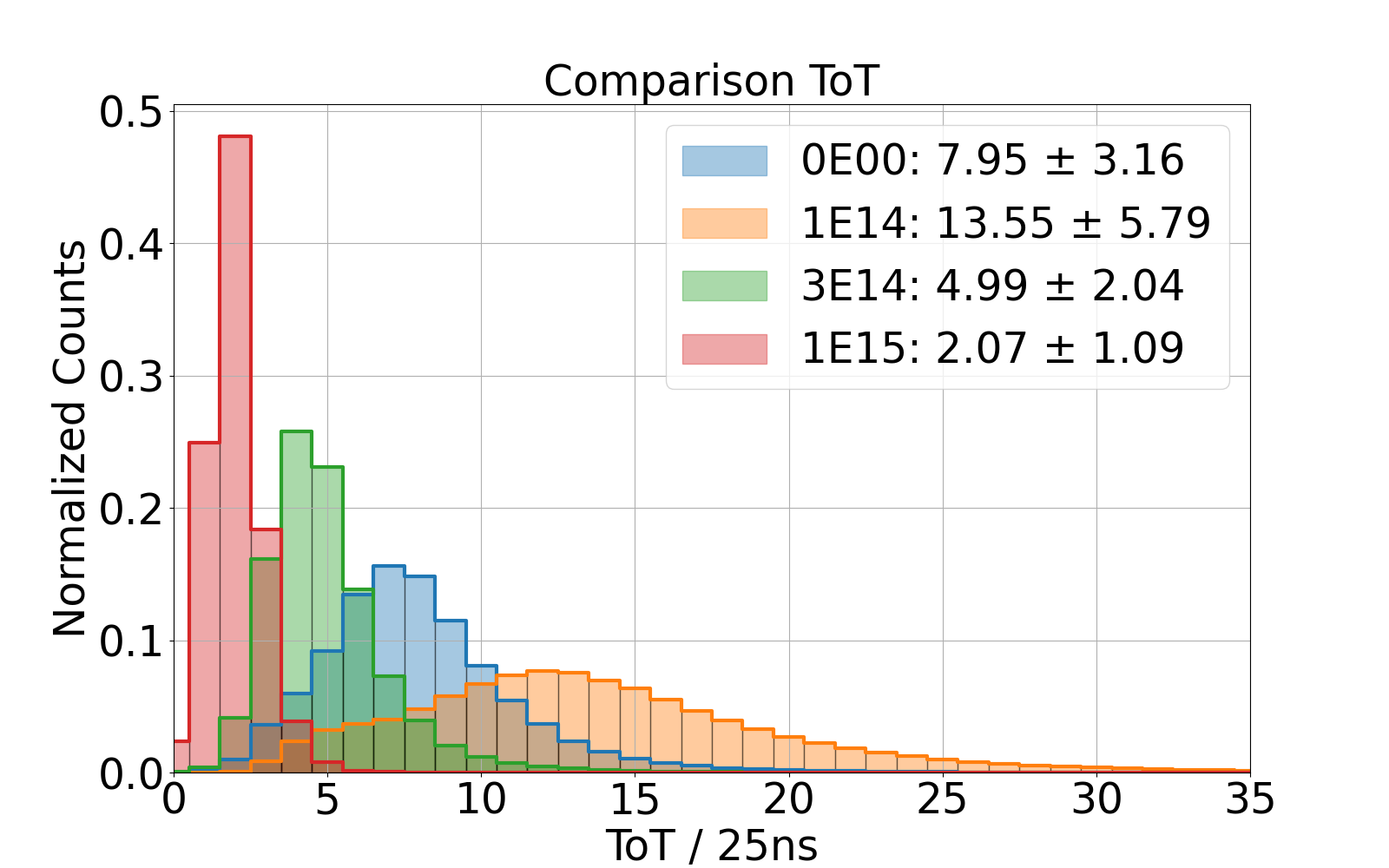}
\caption{Comparison of the mean normalized time over threshold in units of \SI{25}{ns}.}
\label{fig:directCompTot}
\end{subfigure}
\caption{Direct comparison of the various fluence levels recorded with shared threshold and bias voltage settings. The non-irradiated sample is referred to as \emph{0E00} here.}
\label{fig:directComp}
\end{figure}

To enable a direct comparison of the irradiation effects on the \mpw, one operating condition was chosen for the operation of all samples. It consists of  a bias voltage of \SI{190}{V} and a threshold setting of $V_{\text{Thr}} = \SI{200}{mV}$ above baseline, which corresponds to a charge of $Q_{\text{Thr}} \approx \SI{5000}{e^-}$. Such a high threshold was chosen to avoid any problems related to noise. A summary of the results is presented in table~\ref{tab:directComp}. The comparison of the cluster size, defined by the number of directly neighboring pixels that got hit within a time-window of $\SI{20}{\mu s}$,  as well as the time over threshold (ToT), which is also depicted in figure~\ref{fig:directComp}, gives insights about possible increased charge trapping probabilities. As expected, both the cluster size and the ToT distributions shift towards lower values at higher irradiation levels. A comparison of the sample irradiated to $\SI{1E14}{n_{eq}cm^{-2}}$ with the non-irradiated sample shows an increase of both observables. 

As the non-irradiated sensor is already fully depleted at $V_{\text{Bias}}=\SI{190}{V}$, a larger depletion zone due to irradiation is not possible. The increase in cluster size and ToT can instead be attributed to changes in charge sharing, likely caused by modifications of the electric field that introduce lateral components.

\begin{figure}[htbp]
\centering
\begin{subfigure}{0.23\textwidth}
\centering
\includegraphics[width=\textwidth]{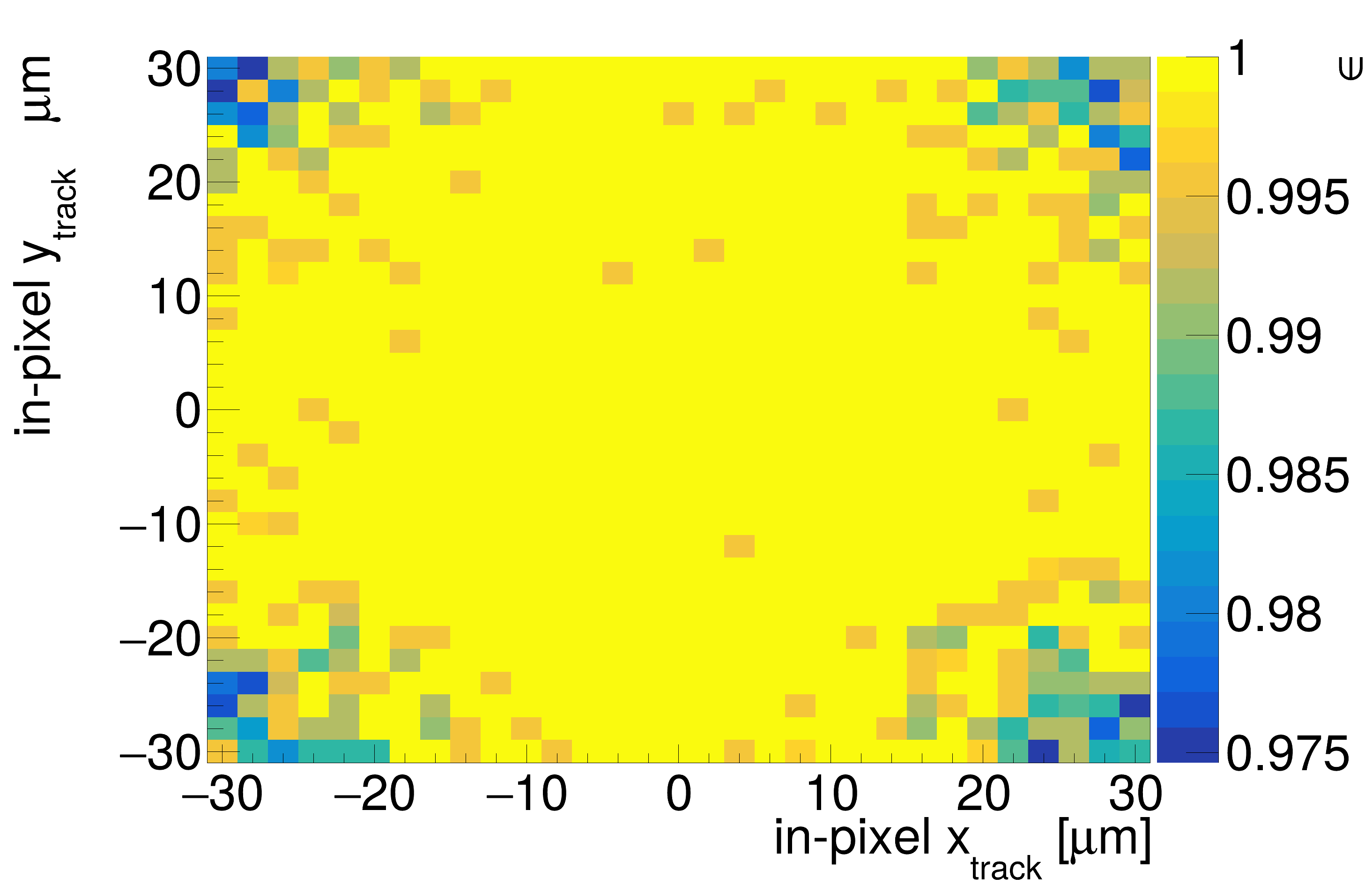}
\caption{In-pixel efficiency of a non-irradiated sample.}
\end{subfigure}
\begin{subfigure}{0.23\textwidth}
\centering
\includegraphics[width=\textwidth]{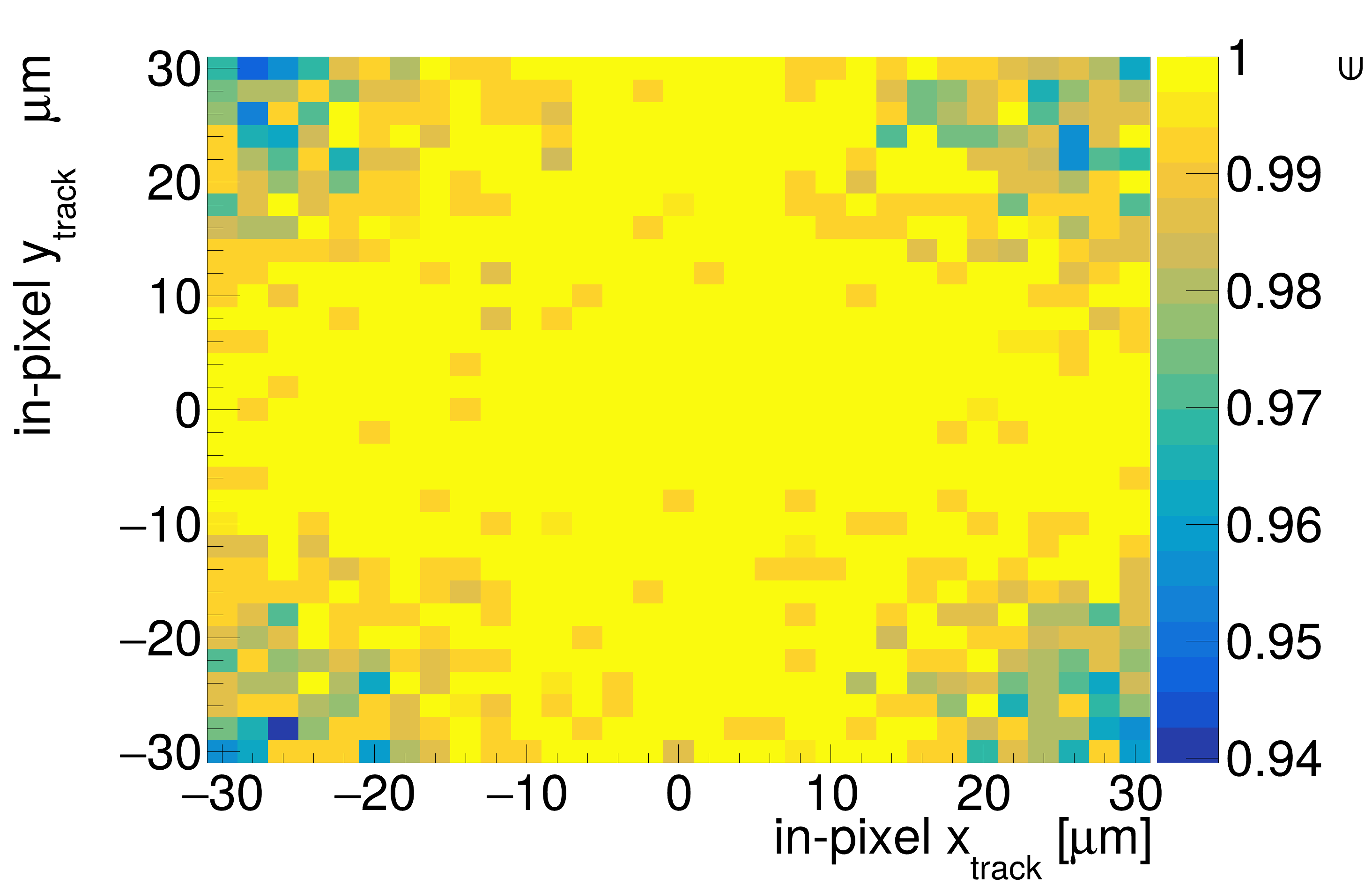}
\caption{In-pixel efficiency of a sample irradiated to $\SI{1E14}{n_{eq}cm^{-2}}$.}
\end{subfigure}
\begin{subfigure}{0.23\textwidth}
\centering
\includegraphics[width=\textwidth]{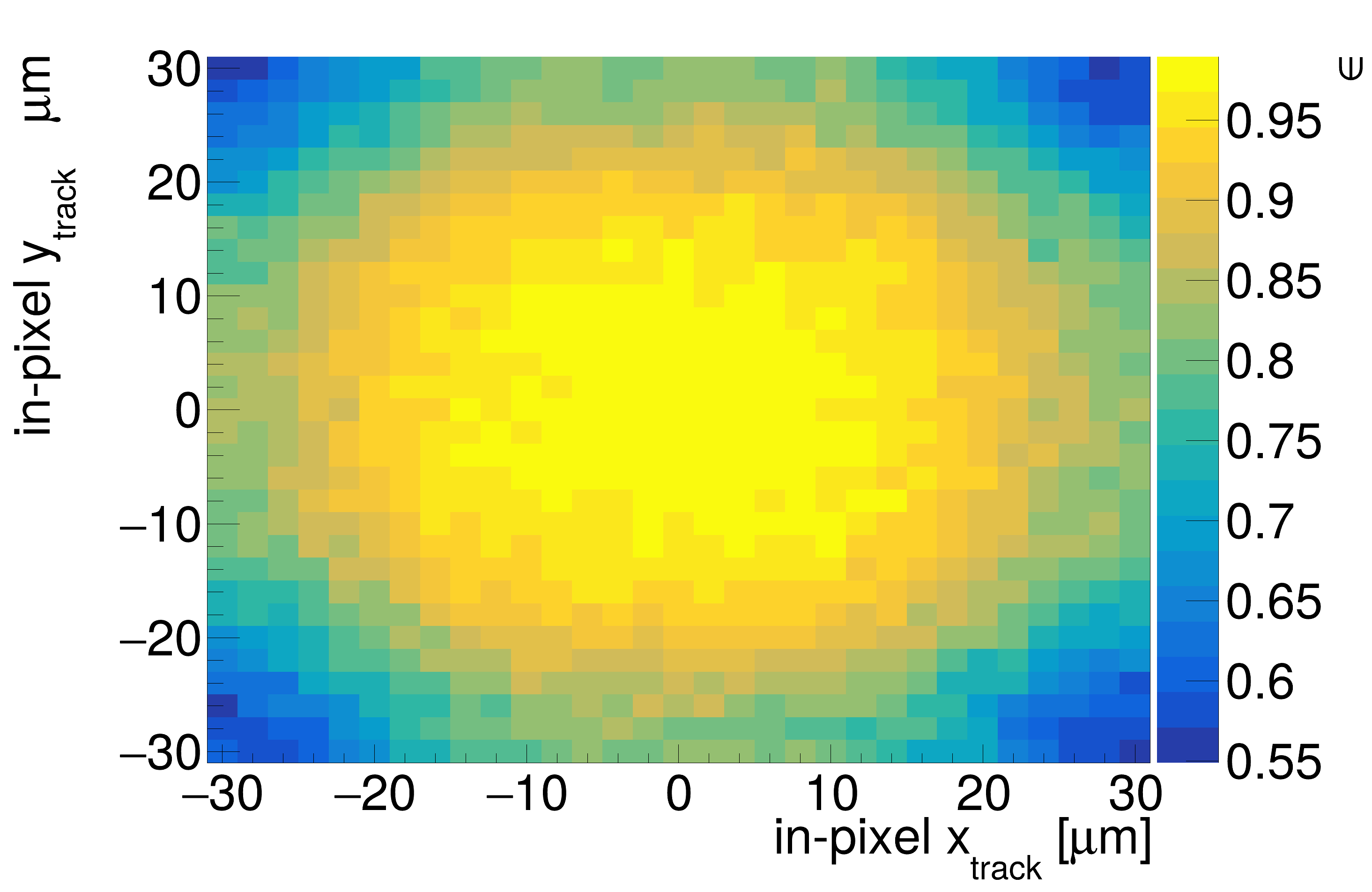}
\caption{In-pixel efficiency of a sample irradiated to $\SI{3E14}{n_{eq}cm^{-2}}$.}
\end{subfigure}
\begin{subfigure}{0.23\textwidth}
\centering
\includegraphics[width=\textwidth]{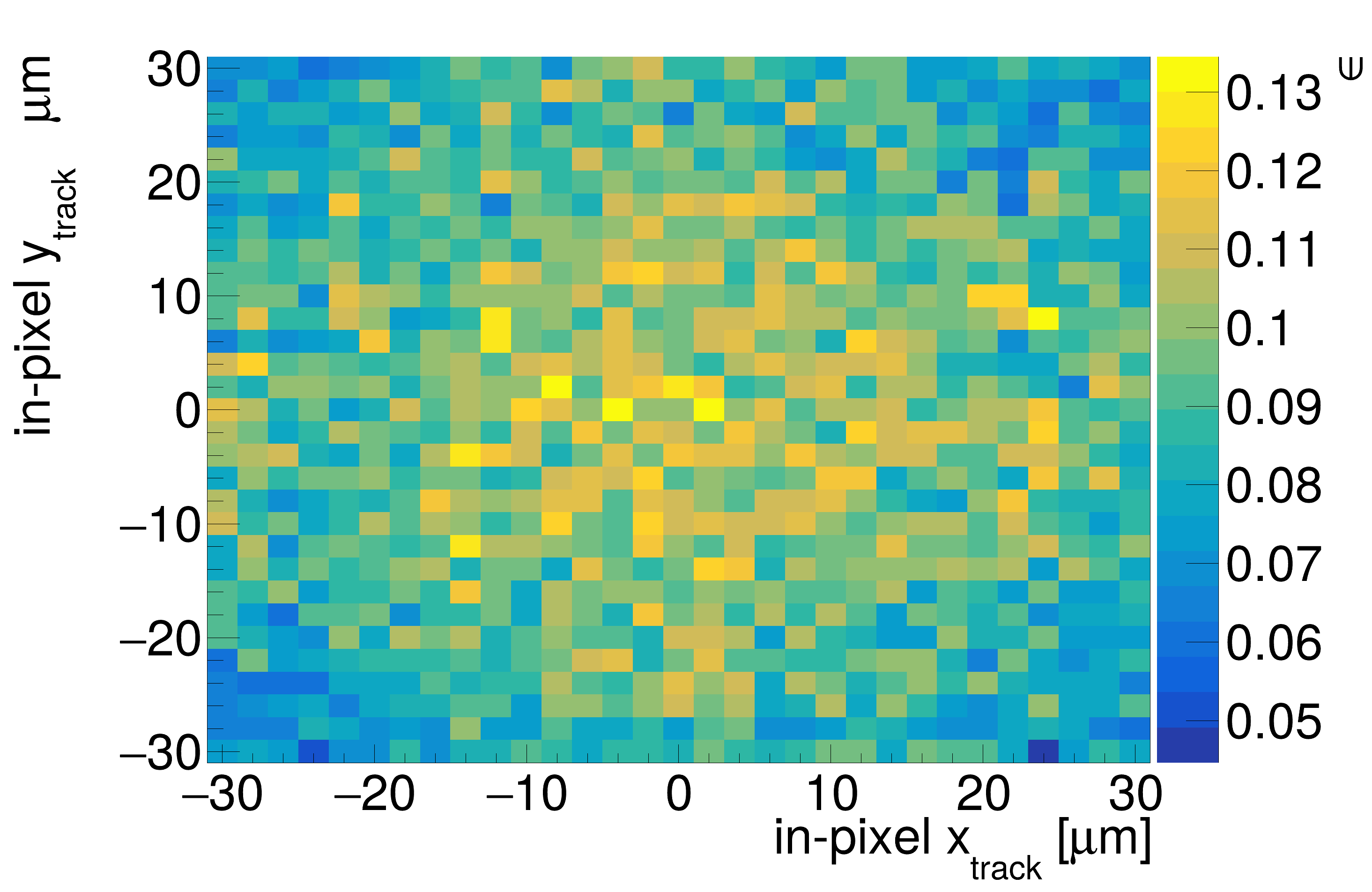}
\caption{In-pixel efficiency of a sample irradiated to $\SI{1E15}{n_{eq}cm^{-2}}$.}
\end{subfigure}
\caption{In-pixel efficiency maps of differently irradiated samples at common bias and threshold settings.}
\label{fig:inPixelEffiStd}
\end{figure}

The hit detection efficiency starts to drop at an irradiation level of $\SI{3E14}{n_{eq}}{cm^{-2}}$ to \SI{85.5}{\%} and at $\SI{1E15}{n_{eq}}{cm^{-2}}$ even to \SI{9.2}{\%}. The reason for these losses can be understood from the in-pixel efficiency maps in figure~\ref{fig:inPixelEffiStd}. Due to the higher charge trapping probability after irradiation, less charge is available in total. In the corners where the charge is shared with neighboring pixels, the available charge is no longer sufficiently large to exceed the high thresholds chosen for these comparison measurements.

\subsection{Utilizing HV to overcome the limits of irradiated samples}

\begin{figure}[htbp]
\centering
\includegraphics[width=.4\textwidth]{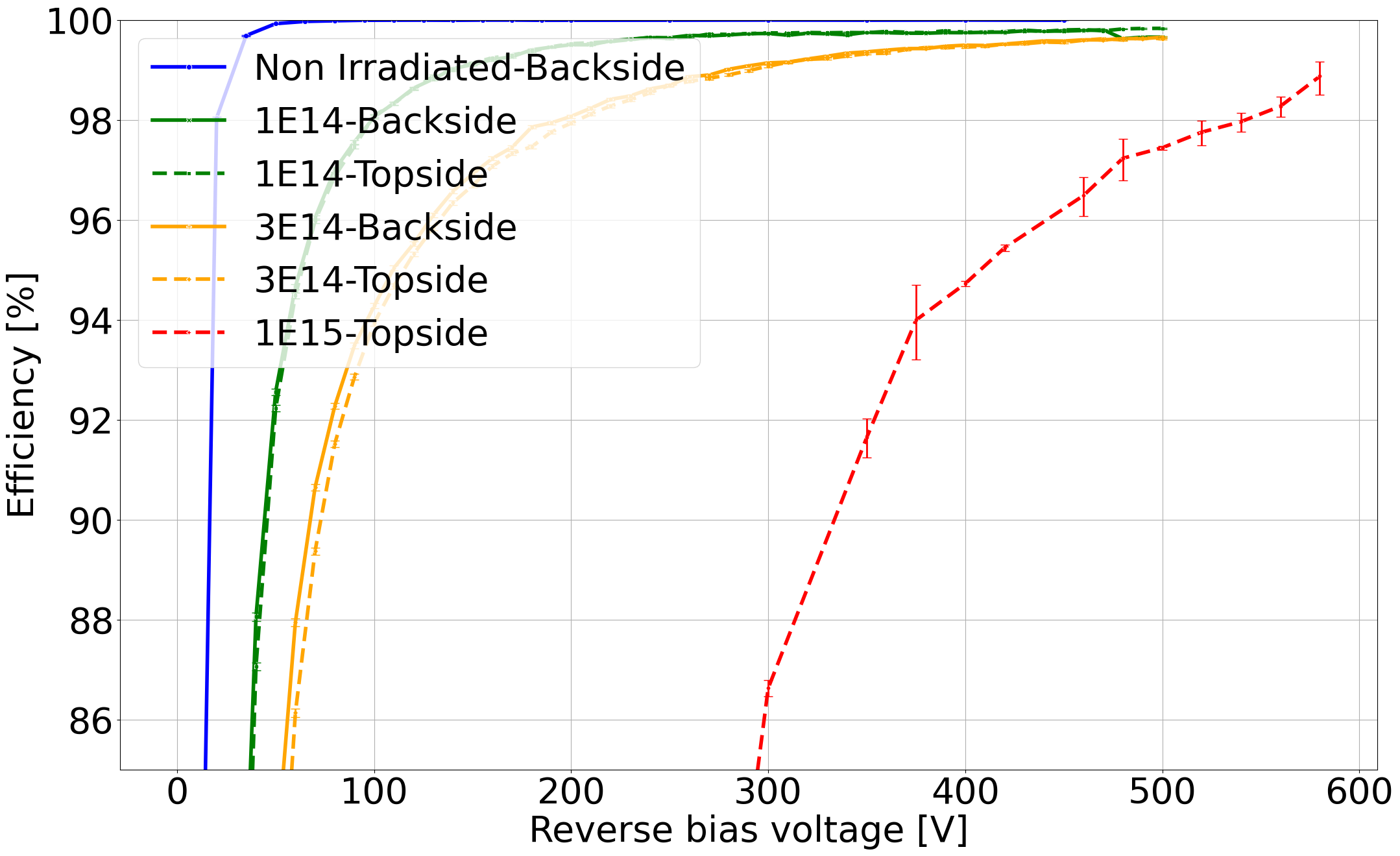}
\caption{Hit detection efficiency as a function of the bias voltage of the various irradiated samples. The prefixes \emph{Topside} and \emph{Backside} in the legend correspond to the applied biasing scheme.}
\label{fig:effiHv}
\end{figure}

Utilizing the biasing capabilities well above \SI{190}{V}, as depicted in figure~\ref{fig:effiHv}, allows to recover the hit detection efficiency to values similar to those pre-irradiation. While the $\SI{1E14}{n_{eq}cm^{-2}}$ sample recovers to $\epsilon > \SI{99}{\%}$ at $V_{\text{Bias}} \approx \SI{150}{V}$, the sample irradiated to $\SI{3E14}{n_{eq}cm^{-2}}$ reaches these $\epsilon$ values at $V_{\text{Bias}} \geq \SI{280}{V}$. The  $\SI{1E15}{n_{eq}cm^{-2}}$ sample manages to recover to $\epsilon \approx \SI{98.9}{\%}$ at $V_{\text{Bias}} \approx \SI{580}{V}$.

\begin{figure}[htbp]
\begin{subfigure}{0.23\textwidth}
\centering
\includegraphics[width=\textwidth]{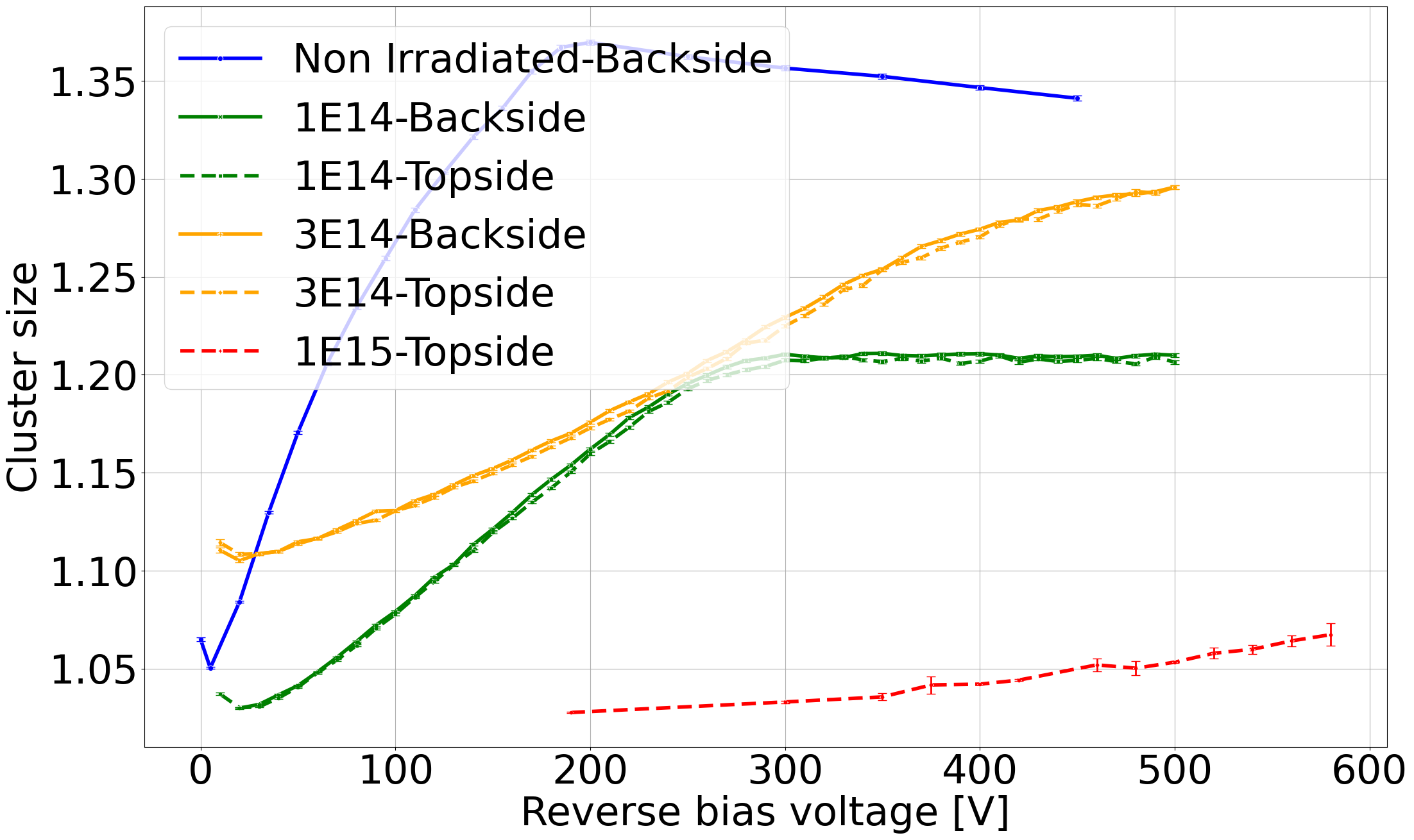}
\caption{Average cluster size.}
\label{fig:clstrsizeHv}
\end{subfigure}
\begin{subfigure}{0.23\textwidth}
\centering
\includegraphics[width=\textwidth]{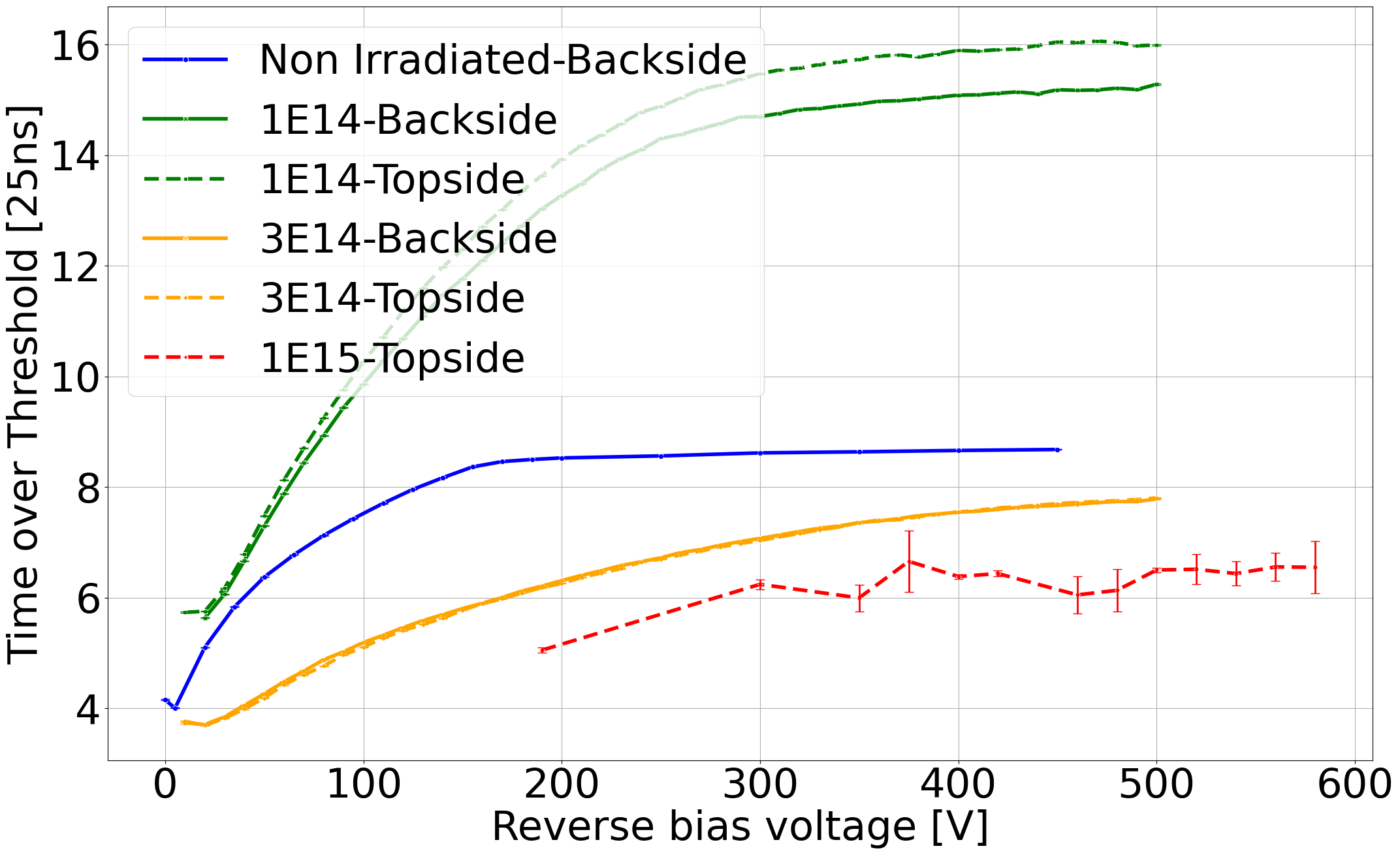}
\caption{Average time over threshold.}
\end{subfigure}
\caption{Charge-related characteristics of the various samples as a function of the bias voltage. Direct comparison due to different threshold settings between the samples is not possible.}
\label{fig:clstrsizeTotHv}
\end{figure}

Investigating saturation effects of the ToT and cluster size as a function of the bias voltage, as presented in figure~\ref{fig:clstrsizeTotHv}, allows to depict if full depletion can be reached. In the case of the non-irradiated sample full depletion is reached at $V_{\text{Bias}} \approx \SI{190}{V}$, while in the case of the $\SI{1E14}{n_{eq}cm^{-2}}$ sample saturation effects are also observed, but at a higher $V_{\text{Bias}} \approx \SI{300}{V}$. The samples irradiated with even higher fluences no longer exhibit these saturation effects, indicating that full depletion can no longer be achieved. It has to be noted that a direct comparison of ToT and cluster size between the different samples can not be made by these observations, as different threshold settings were applied for the various fluence levels.

\begin{figure}[htbp]
\centering
\includegraphics[width=.4\textwidth]{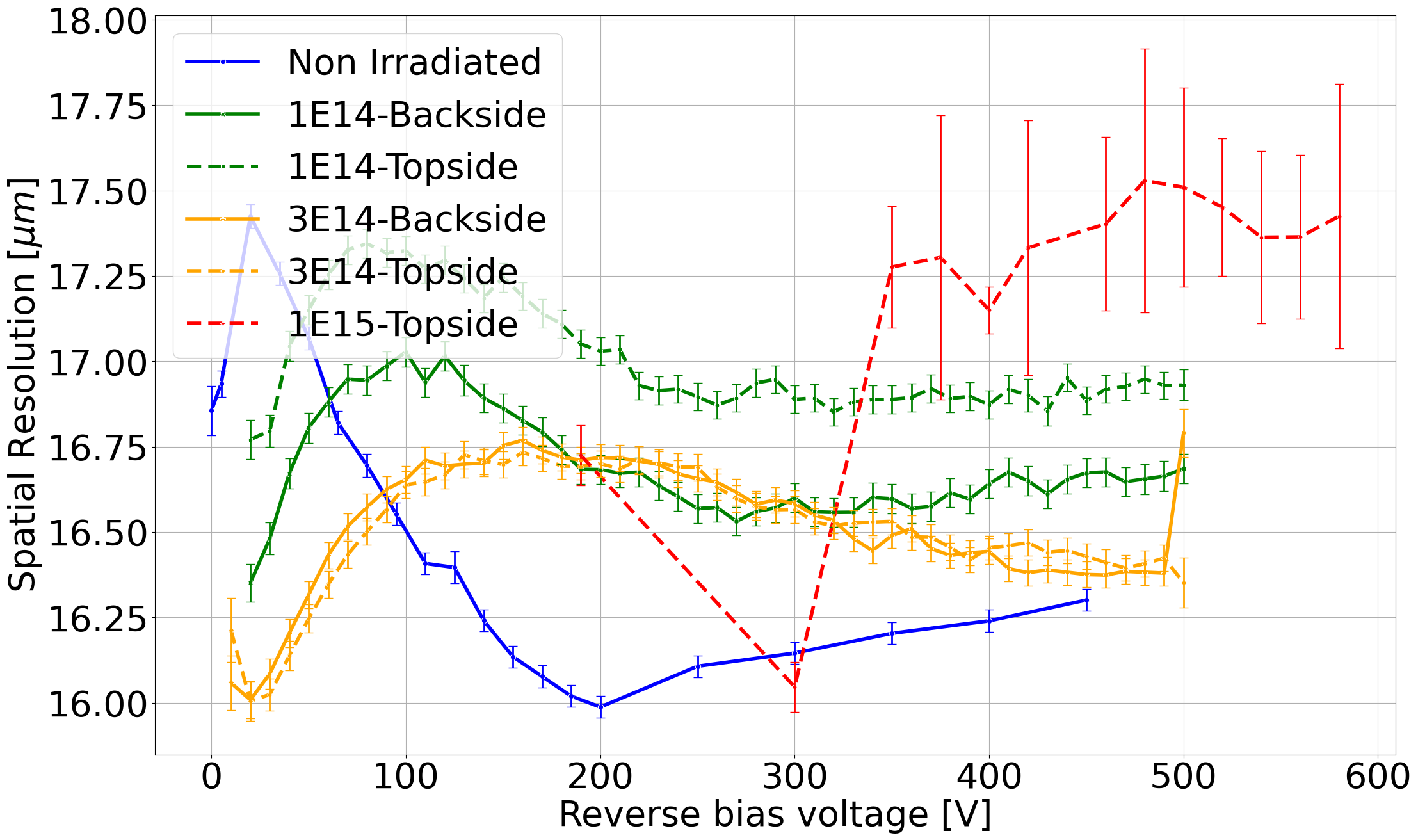}
\caption{Spatial resolution as a function of the bias voltage of the various irradiated samples. The larger error bars for the sample irradiated to $\SI{1E15}{}$ result from  a lack of statistics.}
\label{fig:spatResolution}
\end{figure}

Evaluating the difference between the track intersect at the DUT position $x_{\text{Track}}$ and the cluster center of the detected hits on the DUT $x_{\text{DUT}}$ leads to the so-called spatial residuals $x_{\text{Track}} - x_{\text{DUT}}$. While excluding the DUT in track fitting yields unbiased residuals, including the DUT and explicitly demanding a hit on it for a track to be taken into account leads to biased residuals. The spatial resolution of the various samples at different bias voltages, depicted from the standard deviation of both residual distributions by calculating the geometric mean $\sqrt{\sigma_{\text{Unbiased}} \cdot \sigma_{\text{Biased}}}$ \cite{Alexopoulos_2014}, is presented in figure~\ref{fig:spatResolution}. Comparing these results with the cluster size in figure~\ref{fig:clstrsizeHv} allows us to draw the conclusion that the spatial resolution degrades due to smaller cluster sizes after irradiation. With a mean cluster size $>1$ for all samples, the resolution still exceeds the binary resolution of $\SI{62}{\mu m} / \sqrt{12} \approx \SI{17.9}{\mu m}$. Taking a closer look into the non-irradiated sample, we observe a minimum (best value) of the spatial resolution of $\approx \SI{16}{\mu m}$ at a bias voltage of \SI{200}{V}. At even larger bias voltages, the cluster size starts to decrease from a maximum of $\approx 1.35$ pixel per cluster, which is explainable by the lower drift time of charge carriers in larger electric fields suppressing charge sharing through diffusion.

Across the different irradiation levels, backside biasing showed no significant advantage. Except for the minor difference in the ToT and spatial resolution of the $\SI{1E14}{n_{eq}cm^{-2}}$ sample, no significant distinction is observed between backside and topside biasing. 

\section{Summary and Outlook}

It has been demonstrated that the \mpw can be biased up to several \SI{100}{V} even after irradiation, although the leakage current rises significantly. However, cooling effectively reduces this increase. Additionally, raising the bias voltage allows the restoration of the sensor's detection capabilities to levels comparable to those before irradiation, confirming that the HV-CMOS approach enables the development of radiation-hard detectors.

To investigate the technology's capabilities in terms of radiation hardness even further, another irradiation campaign, targeting the more intermediate fluence levels, with multiple samples irradiated to \SI{5E14}{}, \SI{2E15}{} and \SI{3E15}, was conducted. This time, more samples from a wafer without backside processing were also irradiated to investigate the differences between the two biasing schemes further. The evaluation of this new batch is due, and another test-beam campaign is scheduled for the summer of 2025.

\section{Acknowledgements}
This work has been partly performed in the framework of the CERN-RD50 collaboration.

The measurements leading to these results have partially been performed at the Test Beam Facility at DESY Hamburg (Germany), a member of the Helmholtz Association (HGF).

This project has received funding from the European Union's Horizon Europe Research and Innovation program under Grant Agreement No 101057511 (EURO-LABS).


\bibliographystyle{elsarticle-num-names} 
\bibliography{bibfile}
\end{document}